# Effects of chromatic dispersion on single-photon temporal wave functions in quantum communications

Artur Czerwinski[1,2] · Xiangji Cai[3] · Saeed Haddadi[4]

## Abstract

In this study, we investigate the effects of chromatic dispersion on single-photon temporal wave functions (TWFs) in the context of quantum communications. Departing from classical beam analysis, we focus on the temporal shape of single photons, specifically exploring generalized Gaussian modes. From this foundation, we introduce chirped and unchirped Gaussian TWFs, demonstrating the impact of the chirp parameter in mitigating chromatic dispersion effects. Furthermore, we extend our investigation to time-bin qubits, a topic of ongoing research relevance. By exploring the interplay of dispersion effects on qubit interference patterns, we contribute essential insights to quantum information processing. This comprehensive analysis considers various parameters, introducing a level of complexity not previously explored in the context of dispersion management. We demonstrate the relationships between different quantities and their impact on the spreading of TWFs. Our results not only deepen the theoretical understanding of single-photon TWFs but also offer practical guidelines for system designers to optimize symbol rates in quantum communications.

B Artur Czerwinski
aczerwin@umk.pl

Xiangji Cai
xiangjicai@foxmail.com

Saeed Haddadi
haddadi@semnan.ac.ir

1   Institute of Physics, Faculty of Physics, Astronomy and Informatics, Institute of Physics, Nicolaus Copernicus University in Torun, ul. Grudziadzka 5, 87-100 Torun, Poland

2   Centre for Quantum Optical Technologies, Centre of New Technologies, University of Warsaw, Banacha 2c, 02-097 Warszawa, Poland

3   School of Science, Shandong Jianzhu University, Jinan 250101, China

4   Faculty of Physics, Semnan University, P.O. Box 35195-363, Semnan, Iran

Springer

# 1 Introduction

Quantum communications is a branch of quantum mechanics that deals with the transmission and manipulation of quantum states of light or matter to facilitate the exchange of information [1, 2]. It is an area of active research, with the goal of developing secure, high-speed, and long-distance communication systems based on quantum-mechanical principles [3–5]. Effective manipulation and control of light pulse shapes are essential for these applications [6].

In practice, Gaussian pulses are often used as a model for optical pulses generated by laser sources, as they closely approximate the shape of the pulses generated in these systems [7]. Gaussian pulses are used in a variety of quantum communication protocols, such as continuous-variable quantum key distribution (CV-QKD), quantum teleportation and time-bin encoding [8–10]. In CV-QKD, Gaussian pulses are used to encode the quantum state of a light field, which can then be transmitted over an optical fiber to a remote receiver. The receiver measures the light field and uses the information contained in the Gaussian pulse to establish a shared secret key with the sender [11, 12]. In quantum teleportation, Gaussian pulses are used to transfer the quantum state of a light field from one location to another. This is done by first entangling a pair of light fields, and then transmitting one of the light fields (the "idler") to the receiver. The sender then encodes the quantum state of the light field to be teleported onto the other light field (the "signal"), and sends it to the receiver. The receiver then uses the entanglement to reconstruct the original quantum state of the light field [13, 14]. In both of these protocols, the shape and quality of the Gaussian pulses are critical for achieving high-fidelity transmission of the quantum information. Therefore, techniques such as chirp management and dispersion compensation are often used to preserve the shape of the Gaussian pulses over long distances.

In time-bin encoding, Gaussian wave packets are used to represent the quantum state of a single photon in the time domain, rather than the traditional frequency or spatial domains [15–17]. Exploiting the time-bin degree of freedom allows for encoding quantum information in terms of the relative arrival times of photons. This technique offers a particularly robust type of quantum information [18]. Advantages of time-bin encoding were demonstrated in 1999 by the Group of Applied Physics from the University of Geneva [19]. They generated arbitrary time-bin qubits by sending a single-photon wave packet through a Mach–Zehnder interferometer. Due to a difference in the length of the optical paths, the photon wave packet leaves the interferometer in a quantum-mechanical superposition of "early" and "later" time bins. The Geneva team became renowned for demonstrating the ability of time-bin qubits to travel extended distances through optical fibers with minimal decoherence.

In quantum communications, the light pulse can be affected by a variety of factors that can degrade the quality of the quantum state being transmitted, including chromatic dispersion, noise, photon loss, and misalignment [20]. Chromatic dispersion is a phenomenon that occurs in optical fibers, where different wavelengths of light travel at slightly different speeds [21]. This results in a spreading out of a pulse of light as it propagates through the fiber, which can cause distortion and limit the maximum distance that a signal can travel [22, 23]. Chromatic dispersion depends on a variety of factors, including the properties of the fiber itself and the way that the fiber is manufactured. In general, the destructive effects of chromatic dispersion can be mitigated by using dispersion-compensating fibers or devices [24–26]. However, under some circumstances, the distance of secure communication can be extended by introducing additional dispersion [27].

Chromatic dispersion, an intrinsic phenomenon in optical systems, leads to the temporal spreading of light pulses, which is a well-studied topic in classical optics [28, 29]. However, our study takes a pioneering approach by delving into the realm of quantum mechanics, focusing on the characteristics of single-photon temporal wave functions (TWFs) [30] instead of classical pulses of light. The quantum perspective is essential because emerging technologies, such as quantum key distribution and quantum teleportation, often rely on precise control and manipulation of single-photon states. By investigating how chromatic dispersion affects the temporal modes of single photons, this paper not only enhances our fundamental understanding of the quantum behavior of light but also provides valuable insights for the development of quantum optical systems that can mitigate dispersion-induced limitations, thereby pushing the boundaries of quantum information science and technology.

The paper is organized as follows. Section 2 contains preliminaries related to theoretical background and notations. In Sect. 3, we discuss the broadening of single-photon TWFs represented by generalized Gaussian modes. Based on such a general type of the temporal mode, two specific cases can be distinguished. First, Sect. 4 covers an examination of the effects of chromatic dispersion on Gaussian modes, involving the impact of the chirp parameter. In particular, we investigate the impact of chromatic dispersion on key generation rate in a quantum cryptographic protocol. In Sect. 5, we present a second specific Gaussian model, which involves an analysis of the impact of chromatic dispersion on unchirped Gaussian modes. Moreover, Gaussian modes can be applied to define a time-bin qubit and investigate its properties under chromatic dispersion, which is demonstrated in Sect. 6. We extend the scope of the paper beyond the Gaussian-shaped TWFs by covering hyperbolic-secant modes in Sect. 7. In every model, we consider multiple parameters to present a comprehensive picture of the investigated phenomenon. Finally, in Sect. 8, we present the summary of findings and conclusions.

## 2 Theoretical preliminaries

A single-photon TWF provides a mathematical description of the state of a photon with respect to time. It is a complex-valued function that indicates the probability amplitude of finding the photon at a particular moment in time. Throughout the paper, this object shall be denoted by $\psi(t)$.

The square of the magnitude of the wave function gives the probability density function (PDF) of the existence of the photon at a specific temporal point, i.e., $|\psi(t)|^2 = \psi^*(t)\psi(t) \equiv p(t)$. We consider only the normalized TWFs, which means that $\int_{-\infty}^{\infty} \psi^*(t)\psi(t)\,\mathrm{d}t = 1$. The width of a TWF can be characterized by the standard deviation (SD), denoted by $\sigma$ and obtained as the square root of the variance. We will

compute

$$\sigma = \sqrt{\int_{-\infty}^{\infty} t^2 \, p(t) \, dt}, \quad (1)$$

assuming that the PDF is a zero-mean distribution, i.e., $\int_{-\infty}^{\infty} t \, p(t) \, dt = 0$. The assumption of zero-mean implies that we consider $\psi(t)$ to be symmetric around zero, meaning that the central point of the distribution is zero.

Then, we investigate how the TWF changes as a photon propagates through a dispersive medium, like a fiber or the air in the case of free-space optics (FSO). To mathematically represent the effects of chromatic dispersion, we implement a propagator, $\mathcal{S}(t, \tau, L)$ [31–33], which acts on the initial TWF as

$$\psi_L(t) := \int_{-\infty}^{\infty} \mathcal{S}(t, \tau, L) \, \psi(\tau) \, d\tau. \quad (2)$$

The propagator $\mathcal{S}(t, \tau, L)$ can be represented as

$$\mathcal{S}(t, \tau, L) = \frac{1}{2\sqrt{\pi i \beta L}} \exp\left(\frac{i(t-\tau)^2}{4\beta L}\right), \quad (3)$$

where $L$ denotes the propagation distance and $\beta$ stands for the second-order dispersion parameter of the medium, i.e., the GVD parameter. Analysis of (2) allows one to investigate a change of the shape of the output photon compared to the input.

The second-order dispersion parameter, also known as the group velocity dispersion (GVD) parameter, is a measure of the variation of the group velocity of light with wavelength in an optical fiber. It is typically specified in units of $ps^2$/km and is used to predict the amount of pulse spreading that will occur in a fiber over a given distance. Positive GVD indicates normal dispersion, where longer wavelengths travel slower than shorter wavelengths, and negative GVD indicates anomalous dispersion, where longer wavelengths travel faster than shorter wavelengths. The GVD parameter is important in the design of high-speed optical communication systems, as it determines the maximum bit rate that can be transmitted over a given length of fiber.

For the TWF affected by the propagator, we can compute the PDF, which is denoted by $p_L(t)$. Then, we proceed analogously as in (1) to calculate the width of the distribution after the transmission of the signal ($\sigma_L$). Finally, to quantify the effects of chromatic dispersion, we introduce the broadening parameter defined as

$$\Gamma = \frac{\sigma_L}{\sigma}, \quad (4)$$

which can indicate that the TWF has become either wider (for $\Gamma > 1$) or narrower (for $\Gamma < 1$) than the initial one.

Another figure of merit is introduced in the context of quantum communications. If we assume that every photon carries quantum information encoded, for example, in the photon's polarization, we must determine a detection window related to the symbol duration time. Here, we follow the so-called three-sigma rule, which claims that nearly all values of a random variable lie within three SDs of the mean. Then, the symbol duration time after propagation through a dispersive medium can be defined as $\mathcal{T}_S := 6\sigma_L$. The symbol duration time allows one to compute the symbol rate as $f_S := (\mathcal{T}_S)^{-1}$. The symbol rate is measured in baud (Bd) meaning symbols per second. This figure provides an upper boundary of the attainable gross bit rate in quantum communications with single photons.

**Table 1** GVD parameters at $\lambda = 800$ nm based on Ref. [34]

| Nitrogen | Air | Oxygen | Carbon dioxide |
|---|---|---|---|
| 18.70 $\frac{\text{fs}^2}{\text{m}}$ | 20.05 $\frac{\text{fs}^2}{\text{m}}$ | 24.76 $\frac{\text{fs}^2}{\text{m}}$ | 30.90 $\frac{\text{fs}^2}{\text{m}}$ |

In the paper, we analyze and compare the transmission of photons through selected gases with fiber-based propagation. These two scenarios correspond to positive and negative values of the GVD parameter, respectively. As for free-space transmission, we incorporate from Ref. [34] four values of the second-order dispersion parameter that correspond to selected gases; see Table 1. The dispersion measurements were performed using an ultrabroadband femtosecond oscillator. These data allow for a comparison of different gases in terms of their efficiency in FSO communications.

As for fiber-based transmission, we consider a typical single-mode optical fiber (e.g., SMF-28e+) characterized by $\beta = -1.15 \times 10^{-26}$ s$^2$/m at $\lambda = 1550$ nm. Throughout the paper, when we refer to a specific propagation medium, we imply that the corresponding GVD parameter takes the value as provided in this section.

## 3 Broadening of generalized Gaussian modes

### 3.1 Methods

Let us start with a framework for photons characterized by generalized Gaussian temporal modes. Compared to the standard Gaussian distribution, the generalized Gaussian distribution (GGD) is used as an alternative because only an additional shape parameter estimate is required, which is widely used to describe the statistical properties of classical and quantum signals [35, 36]. In this case, we use the properties of GGD and represent the TWF as

$$\psi(t) = \sqrt{\frac{q}{2\alpha\Gamma(1/q)}} \exp\left(-\frac{1+iC}{2\alpha^q}|t|^q\right), \qquad (5)$$

where $\alpha = \sigma\sqrt{\Gamma(1/q)/\Gamma(3/q)}$ with the Gamma function $\Gamma(s) = \int_0^\infty x^{s-1}e^{-x}\mathrm{d}x$ ($s > 0$), $\sigma$ representing the SD of PDF, and the shape parameter $q > 0$. The symbol $C$ denotes a chirp parameter, which is related to the phase of the temporal mode, cf. [37]. For $q = 1/2$, $q = 1$, and $q = 2$, the GGD represents the Gamma distribution, Laplacian distribution, and Gaussian distribution, respectively. The case of Gaussian distribution will be studied thoroughly in Sects. 4 and 5. In the following, we mainly focus on the cases of sub-Gaussian ($0 < q < 2$) and sup-Gaussian ($q > 2$) distributions and make a comparative study of the results with that of the Gaussian case.

The TWF after propagation can be formally expressed in terms of the propagator in (3) acting on the initial representation (5) as

$$\psi_L(t) = \sqrt{\frac{q}{8\pi i \alpha \beta L \Gamma(1/q)}} \exp\left(\frac{it^2}{4\beta L}\right) \\ \times \int_{-\infty}^{\infty} \exp\left(-\frac{it}{2\beta L}\tau + \frac{i}{4\beta L}\tau^2 - \frac{1+iC}{2\alpha^q}|\tau|^q\right) d\tau. \quad (6)$$

The relative ratio $\Gamma = \sigma_L/\sigma$ is used to quantify the broadening of generalized Gaussian modes. For the case of Gaussian mode $q = 2$, the ratio $\Gamma$ can be written in the analytical form as studied in Sect. 4.1. However, for the general cases of sub- and sup-Gaussian modes, we can obtain it only numerically and follow it versus the transmission distance $L$.

### 3.2 Results and analysis

In this section, we consider the impact of chromatic dispersion on the broadening of generalized Gaussian modes, characterized by the shape parameter $q$. The quantification of this impact is achieved through the computation of the broadening parameter $\Gamma$.

In this analysis, let us assume the initial SD equals $\sigma = 4.25$ ps. As for the dispersive medium, we compare fiber-based propagation with FSO transmission through the air.

The observed trends in the broadening parameter for generalized Gaussian modes are intriguing and offer significant insights. Remarkably, for non-zero chirp parameters, the parameter does not exhibit a monotonous behavior with increasing propagation distance. Instead, it follows a pattern, characterized by an initial decrease, reaching a minimum, and subsequently an increase.

In Fig. 1, two distinct cases are considered, namely $q = 1$ (Fig. 1a) and $q = 8$ (Fig. 1b), assuming the transmission through the air. These figures provide valuable insight into how the shape parameter influences the broadening behavior. Notably, for $q = 1$, the broadening parameter experiences a deeper decline compared to $q = 8$. This suggests that the initial mode shape significantly affects how chromatic dispersion impacts the TWF, with $q = 1$ demonstrating a more pronounced mitigation of broadening.

In Fig. 2, a similar scenario is depicted, but the transmission through a typical SMF fiber is considered. To achieve a decline of the broadening parameter, we allow for negative values of the chirp parameter. Due to more significant dispersion, the range of $L$ for which the broadening parameter drops below one is much smaller than in the case of Fig. 1.

In Figs. 1 and 2, within each $q$, the impact of the chirp parameter is also investigated. The results reveal a direct relationship between the chirp and broadening parameters. Specifically, higher values of $|C|$ are associated with a more profound decline in the broadening parameter. This suggests that effective management of chirp parameters can play a crucial role in mitigating the detrimental effects of chromatic dispersion on generalized Gaussian modes.

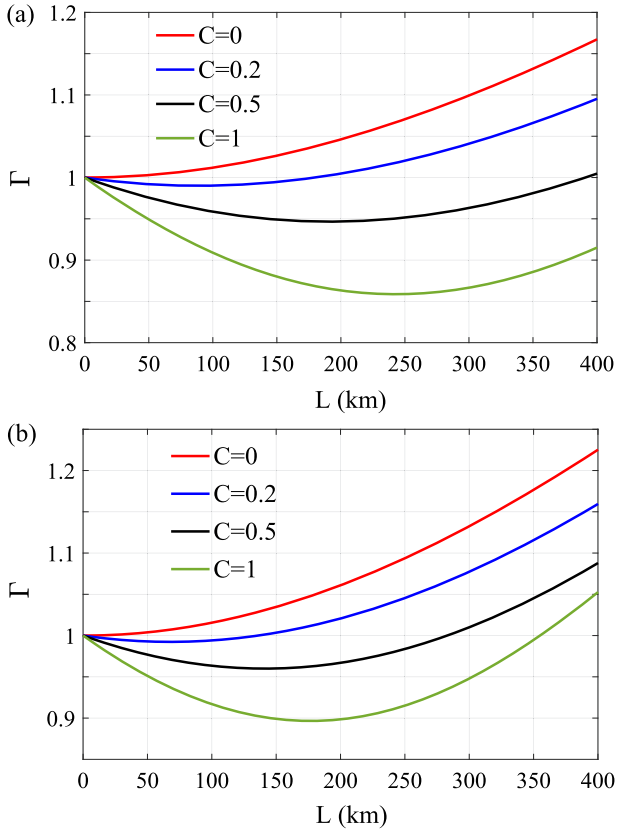

**Fig. 1** The broadening parameter, $\Gamma$, for $L \in [0, 400]$ km, $\sigma = 4.25$ ps and four values of $C$ for **a** the sub-Gaussian mode $q = 1$ and **b** the sup-Gaussian mode $q = 8$. The transmission through the air is considered (color figure online)

Moreover, in Figs. 3 and 4, we present the broadening parameter for various values of $q$, considering the transmission through the air and an SMF-28e+ fiber, respectively. The results imply that certain shape parameters may offer superior resistance to chromatic dispersion, allowing for the maintenance of a narrower TWF over extended propagation distances.

### 3.3 Applications in quantum communications

The results on generalized Gaussian modes carry significant implications for the design and implementation of quantum communication systems. Understanding the interplay between shape parameters, chirp values, and chromatic dispersion allows for

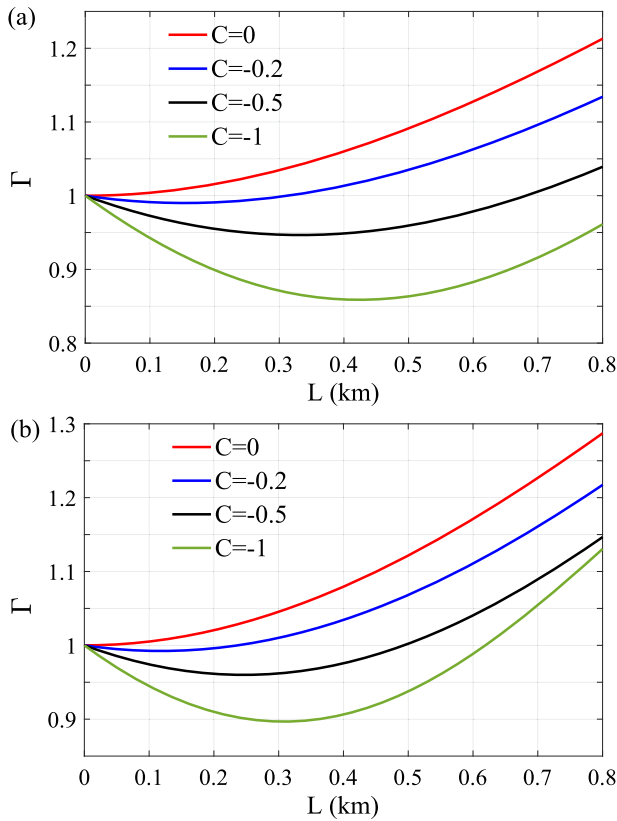

**Fig. 2** The broadening parameter, $\Gamma$, for $L \in [0, 0.8]$ km, $\sigma = 4.25$ps and four values of $C$ for **a** the sub-Gaussian mode $q = 1$ and **b** the sup-Gaussian mode $q = 8$. The transmission through an SMF-28e+ fiber is considered (color figure online)

the optimization of system parameters to achieve high-fidelity transmission over long distances.

In this context, we compute the symbol rate, which is presented in Fig. 5. We consider only fiber-based quantum communication with two values of the chirp parameter: positive ($C = 2$) and negative ($C = -2$).

In the case of positive chirp, presented in Fig. 5a, the symbol rate demonstrates a monotonous decrease with increasing propagation distance. This behavior suggests a consistent broadening of the temporal mode over the distance, reflecting the challenges posed by chromatic dispersion in maintaining a high symbol rate. Conversely, for the negative chirp, depicted in Fig. 5b, the symbol rate exhibits a distinct pattern. Initially, there is an increase in the symbol rate, corresponding to a decline in the temporal width as discussed in the previous section. This initial increase implies a potential advantage in terms of information transfer over short distances. By properly tuning the chirp parameter, one can optimize the performance of the quantum communication system,

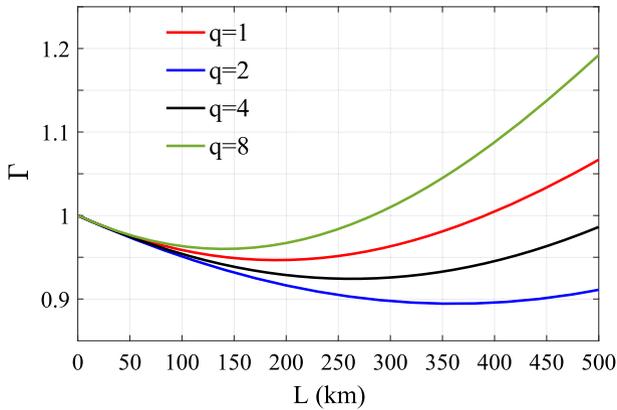

**Fig. 3** The broadening parameter, $\Gamma$, for $L \in [0, 500]$ km, $\sigma = 4.25$ ps, $C = 0.5$ and four values of $q$. The transmission through the air is considered (color figure online)

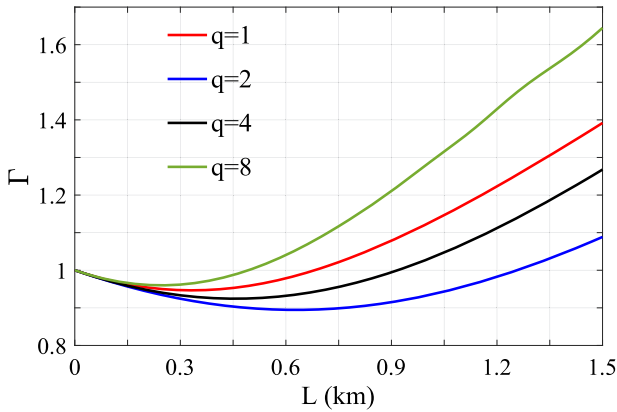

**Fig. 4** The broadening parameter, $\Gamma$, for $L \in [0, 1.5]$ km, $\sigma = 4.25$ ps, $C = -0.5$ and four values of $q$. The transmission through an SMF-28e+ fiber is considered (color figure online)

ensuring high-speed transmission of information and efficient use of the channel's capacity.

The response of the symbol rate to negative chirp is found to be different for various shape parameters. This observation demonstrates the sensitivity of quantum communication systems to the intrinsic characteristics of the temporal modes. The variability in the symbol rate's increase suggests that certain shape parameters may allow for more efficient manipulation of chirp-induced effects, leading to improved symbol rates over specific propagation distances.

The research suggests that careful consideration of chirp and shape parameters can be leveraged to optimize symbol rates in quantum communication. The transient increase observed under negative chirp conditions indicates a potential sweet spot for information transfer. Exploring and exploiting this behavior could pave the way for the development of adaptive quantum communication systems capable of dynamically adjusting to varying propagation conditions.

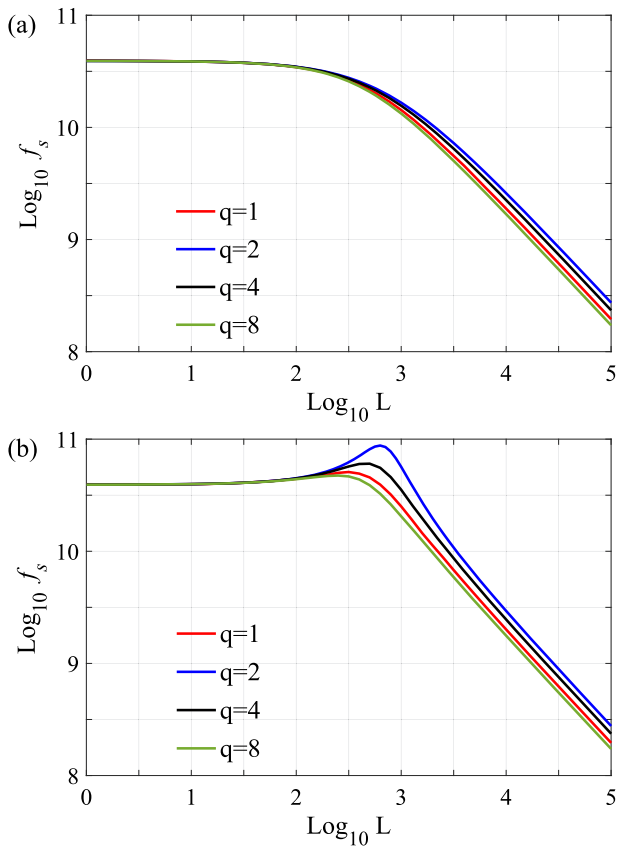

**Fig. 5** Symbol rate, $f_S$, for transmission of photons with generalized Gaussian modes for four values of shape parameter $q$ with **a** positive chirp parameter $C = 2$ and **b** negative chirp parameter $C = -2$ through an SMF-28e+ fiber. Logarithmic scale is used for $L$ expressed in meters (color figure online)

## 4 Broadening of chirped Gaussian modes

### 4.1 Methods

Let us consider a temporal mode of a single photon that is Gaussian and can be described by a following TWF

$$\psi(t) = \frac{1}{\sqrt{\sqrt{2\pi}\sqrt{\sigma}}} \exp\left(-\frac{1+iC}{4\sigma^2}t^2\right), \qquad (7)$$

As indicated in Sect. 3, this type of TWF is a special case of the generalized Gaussian corresponding to the shape parameter $q = 2$. Nevertheless, chirped Gaussian modes deserve a separate treatment and in-depth analysis due to their significance for laser physics.

By using the propagator given in Eq. (3), we arrive at the formula to describe the TWF after propagation

$$\psi_L(t) = \frac{(1-i)\exp\left(\frac{1+iC}{4(C-i)L\beta - 4\sigma^2}t^2\right)}{2^{3/4}\pi^{1/4}\sqrt{\frac{L\beta(1+iC)-i\sigma^2}{\sigma}}} \quad \text{for } \beta > 0 \tag{8}$$

and

$$\psi_L(t) = \frac{(-2\pi)^{-1/4}\exp\left(\frac{1+iC}{4(C-i)L\beta - 4\sigma^2}t^2\right)}{\sqrt{\frac{i[(i-C)L\beta+\sigma^2)]}{\sigma}}} \quad \text{for } \beta < 0. \tag{9}$$

Then, we compute the SD of the modified PDF function. We obtain

$$\sigma_L = \frac{\sqrt{L^2\beta^2 + (\sigma^2 - CL\beta)^2}}{\sigma}, \tag{10}$$

for both $\beta > 0$ and $\beta < 0$.

Finally, we get the relative ratio $\sigma_L/\sigma$ to quantify the broadening of chirped Gaussian modes

$$\Gamma = \frac{\sqrt{L^2\beta^2 + (\sigma^2 - CL\beta)^2}}{\sigma^2} = \sqrt{\left(1 - \frac{CL\beta}{\sigma^2}\right)^2 + \left(\frac{L\beta}{\sigma^2}\right)^2}. \tag{11}$$

### 4.2 Results and analysis

We notice that when $\beta > 0$, we obtain an initial decrease in the photon's width for every $C > 0$. This tendency is depicted in Fig. 6, where the transmission through the air is considered for selected values of the chirp parameter. The initial width parameter was the same as in Sect. 3, i.e., $\sigma = 4.25$ ps.

One can observe that the higher the value of $C$, the more significant decline in the width we obtain. To quantify this phenomenon, we solve

$$\frac{d\Gamma}{dL} = 0,$$

which leads to

$$L_{\min} = \frac{C\sigma^2}{(1+C^2)\beta}. \tag{12}$$

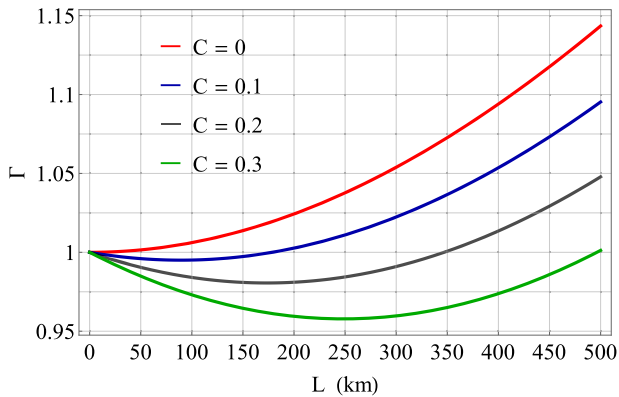

**Fig. 6** The broadening parameter, $\Gamma$, for $L \in [0, 500]$ km and four values of $C$. The transmission through the air is considered (color figure online)

The formula (12) provides the transmission distance that corresponds to the minimal value of the temporal width, which is

$$\sigma_L^{\min} = \frac{\sigma}{\sqrt{1+C^2}}. \tag{13}$$

For $C = 0.3$, we obtain $L_{\min} \approx 250$ km, which corresponds to $\Gamma \approx 0.958$. Moreover, when $L = 500$ km, then $\Gamma \approx 1$, which means that the pulse regains its initial width.

Furthermore, dispersion in an SMF-28e+ fiber can be investigated. In such a case, negative chirp parameters have to be incorporated to witness a decrease in the temporal width. In Fig. 7, the results are presented. One can immediately notice that the overall impact of dispersion on the temporal shape of photons is much more significant than in the case of Fig. 6. Nevertheless, up to some distance, one can still control the broadening by adjusting the chirp parameter so that the pulse spreading is not substantial.

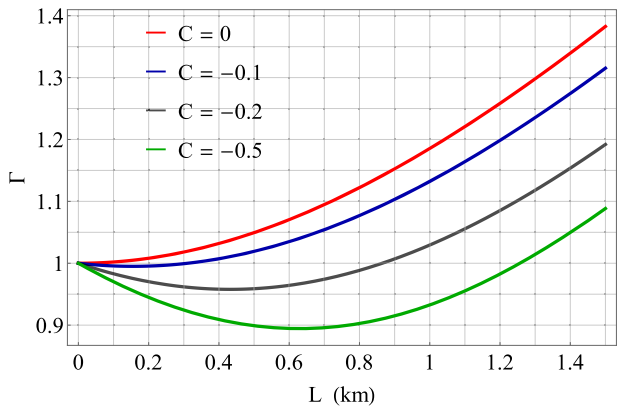

**Fig. 7** The broadening parameter, $\Gamma$, for $L \in [0, 1.5]$ km and four values of $C$. The transmission through an SMF-28e+ fiber is considered (color figure online)

### 4.3 Applications in quantum communications: symbol rate

The results presented in Sect. 4.2 focused on the fact that for chirped Gaussian TWF, one can modulate the phase $C$ so that the broadening parameter initially declines toward its minimum value. Now, let us analyze what happens if we consider long-distance quantum communications. In this context, we again use the concept of symbol rate given by

$$f_S = \left(6\frac{\sqrt{L^2\beta^2 + (\sigma^2 - CL\beta)^2}}{\sigma}\right)^{-1}. \tag{14}$$

In the logarithmic scale, we compute $f_S$ up to $L = 10{,}000$ km, which may correspond to the length of intercontinental fiber-based links. We consider two scenarios—the positive (in Fig. 8) and the negative (in Fig. 9) values of the chirp parameter.

From Fig. 8, when $C\beta < 0$, it can be seen that the decreasing trend of the symbol rate depends not only on the increase of the transmission distance $L$ but also on the values of the chirp parameter $C$. As expected, for short transmission lengths (up to $L \approx 10$ m), the effect of the chirp parameter on $f_S$ is insignificant. However, the symbol rate decreases more rapidly for higher values of $C$ over long transmission distances.

For a comparative study between the effects of positive and negative chirp parameters on $f_S$, we draw Fig. 9 for $C\beta > 0$. The general behavior of the curves in Fig. 9 is the same as Fig. 8 for large values of $L$. However, for $L \lesssim 1$ km, we find that the symbol rate boosts but experiences a sudden decline afterwards. As $|C|$ increases, the downward tendency of $f_S$ is more significant. At the same time, for the greatest absolute value of the chirp parameter, i.e., $|C| = 15$, we observe the most meaningful boost in the symbol rate, but it lasts only up to $L = 100$ m.

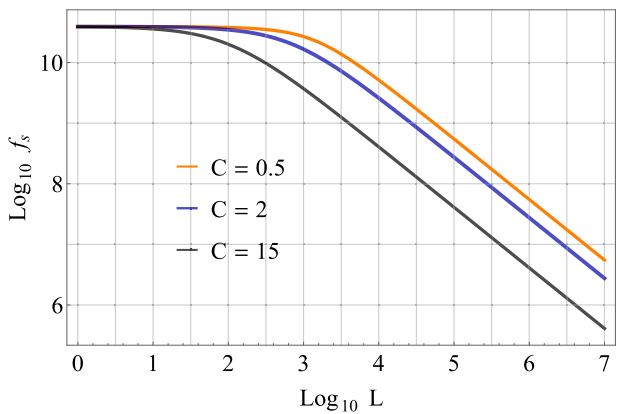

**Fig. 8** Symbol rate, $f_S$, for transmission of photons with Gaussian modes with positive chirp parameters through an SMF-28e+ fiber. Logarithmic scale is used for $L$ expressed in meters (color figure online)

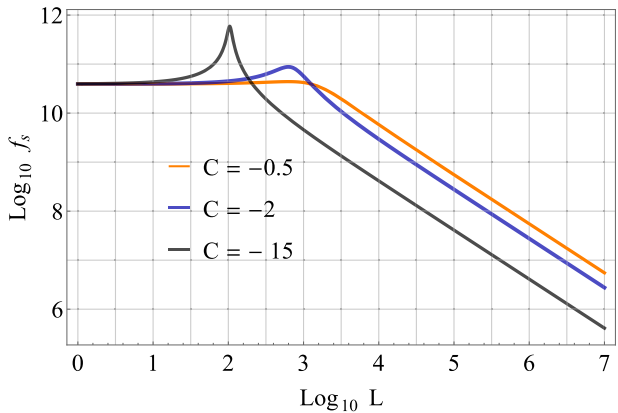

**Fig. 9** Symbol rate, $f_S$, for transmission of photons with Gaussian modes with negative chirp parameters through an SMF-28e+ fiber. Logarithmic scale is used for $L$ expressed in meters (color figure online)

Therefore, the negative values of the chirp parameter in short transmission distances have a different effect on the symbol rate, which causes a significant change in the shape and invalidates the monotonic behavior of the curves compared to Fig. 8. Note that the curves of the function $f_S$ in Fig. 9 decline more sharply than those in Fig. 8 for greater $|C|$.

For a deeper understanding of the effects of the chirp parameter and transmission distance on the symbol rate, we rewrite expression (14) when $C$ and $L$ tend to infinity, namely

$$f_S(C)\,|_{C\to\pm\infty} \approx \frac{\sigma}{6C\sqrt{L^2\beta^2}} + \frac{\sigma^3\sqrt{L^2\beta^2}}{6C^2(L\beta)^3} - \frac{\sqrt{L^2\beta^2}\left(L^2\beta^2\sigma - 2\sigma^5\right)}{12C^3(L\beta)^4} \quad (15)$$

and

$$f_S(L)\,|_{L\to\infty} \approx \frac{\sigma}{6L\sqrt{\beta^2\left(C^2+1\right)}} + \frac{\beta C\sigma^3}{6L^2\left[\beta^2\left(C^2+1\right)\right]^{3/2}} + \frac{\beta^2\left(2C^2-1\right)\sigma^5}{12L^3\left[\beta^2\left(C^2+1\right)\right]^{5/2}}. \quad (16)$$

According to Eqs. (15) and (16), we find that $L$ is a dominant parameter in reducing the symbol rate, and the large values of the chirp parameter (both positive and negative) have destructive effects on it at large distances. These results are in good agreement with Figs. 8 and 9.

### 4.4 Applications in quantum communications: key generation rate

The properties of chirped Gaussian TWFs can be investigated in the context of the BB84 QKD protocol [38]. According to BB84, the information bits 0 and 1 are encoded in the polarization degree of freedom by choosing randomly an orthonormal basis from two possibilities: $\{|H\rangle, |V\rangle\}$ or $\{|D\rangle, |A\rangle\}$. When a sequence of photons is sent from Alice to Bob, the recipient needs to consider the temporal width of each photon to properly register the signal. Let us assume that Bob uses a detector characterized by a Gaussian profile [32, 39]

$$\xi_d(t) = \frac{1}{\sqrt{2\pi\sigma_d^2}} \exp\left(-\frac{t^2}{2\sigma_d^2}\right), \tag{17}$$

where $\sigma_d$ stands for the detector's uncertainty (timing jitter). Then, the PDF of photon detection can be defined by a convolution

$$(p_L \star \xi_d)(t) := \int_{-\infty}^{\infty} p_L(\tau)\xi_d(t-\tau)d\tau \equiv \widetilde{p}_L(t), \tag{18}$$

which allows us to write a formula for the probability of measuring a signal photon within a detection window $w$

$$p_{\text{sig}} = \int_{-w/2}^{w/2} \widetilde{p}_L(t)dt. \tag{19}$$

Moreover, we need to include errors that arise from a possible situation in which Bob detects a photon that precedes or follows the signal. We assume that the temporal separation between consecutive photons is denoted by $\mathfrak{S}$, which gives us two PDFs for false detection

$$\widetilde{p}_L^{\pm}(t) = \int_{-\infty}^{\infty} p_L(\tau \pm \mathfrak{S})\xi_d(t-\tau)d\tau, \tag{20}$$

where '+' refers to a photon following the signal and '−' can be used for the preceding photon. Overall, the probability of registering a wrong photon reads

$$p_{\text{error}} = \int_{-w/2}^{w/2} \widetilde{p}_L^{+}(t)dt + \int_{-w/2}^{w/2} \widetilde{p}_L^{-}(t)dt - 2 \int_{-w/2}^{w/2} \widetilde{p}_L^{+}(t)dt \int_{-w/2}^{w/2} \widetilde{p}_L^{-}(t)dt, \tag{21}$$

where the subtraction of the joint probability is justified by the fact that in the case of two-photon detection in one time slot, Bob is able to recognize a problem and discard such data so it does not affect the key rate.

Next, we propose the following formula to compute the probability of generating a raw key bit

$$p_{\text{raw}} = \frac{\eta(L)[p_{\text{sig}} + p_{\text{error}}(1 - \eta(L)p_{\text{sig}})]}{2}, \quad (22)$$

where $\eta(L) = 10^{-\tilde{\alpha}L}$ represents the channel transmittance for an attenuation coefficient $\tilde{\alpha}$. Eq. (22) represents the overall probability that Bob will obtain a detection during the time window $w$ by randomly choosing a polarization basis for the measurement. Consequently, the quantum bit error rate (QBER) is given by

$$Q = \frac{1}{4} \cdot \frac{p_{\text{error}}(1 - \eta(L)p_{\text{sig}})}{p_{\text{raw}}}. \quad (23)$$

Finally, the key generation rate for BB84 can be computed as [4]

$$\mathcal{K} = \max\{p_{raw}\left(1 - 2h(Q)\right), 0\}, \quad (24)$$

where $h(Q)$ denotes the binary entropy for QBER, i.e., $h(Q) = -Q \log_2 Q - (1 - Q) \log_2(1 - Q)$.

The QKD model presented in this section accounts for both photon loss due to attenuation and errors caused by wrong photon detection. The formula (24) provides the upper bound of the attainable key rate per one signal photon.

In the analysis of the key generation rate, we assume that Bob's detector has perfect efficiency and the system does not detect any dark counts. For the jitter, we assume $\sigma_d = 5$ ps and the initial width of a photon is $\sigma = 4.25$ ps. As for the separation between consecutive photons, we assume that Alice's source generates signal photons with repetition rate 10 GHz, which corresponds to $\mathfrak{S} = 10^{-10}$ s. This value is fixed in our analysis. Photon loss due to attenuation is computed by substituting $\tilde{\alpha} = 0.2$ dB/km, which is a typical attenuation coefficient for an SMF-28e+ fiber corresponding to wavelength $\lambda = 1550$ nm.

From the model, we can expect an interplay between the key generation rate $\mathcal{K}$ and the detection window $w$. From Fig. 10, one can see that $w$ cannot be too low (e.g., $w = 5$ ps) because it decreases the probability of detecting a signal photon. It cannot also be too large (e.g., $w = 150$ ps) because it increases the error probability. Thus, the optimal value is somewhere in between. Hence, we choose $w = 50$ ps as a sort of optimal value to generate key rates presented in Figs. 11 and 12. In Fig. 11, we observe how positive chirp parameters ($C = 1, \ldots, 4$) affect the key generation rate compared to the unchirped case. On the other hand, in Fig. 12, we reveal that negative chirp parameters can be used to increase the key rate. This phenomenon demonstrates that by controlling the value of the chirp parameter we can mitigate the unwanted impact of chromatic dispersion on QKD and increase the key rate.

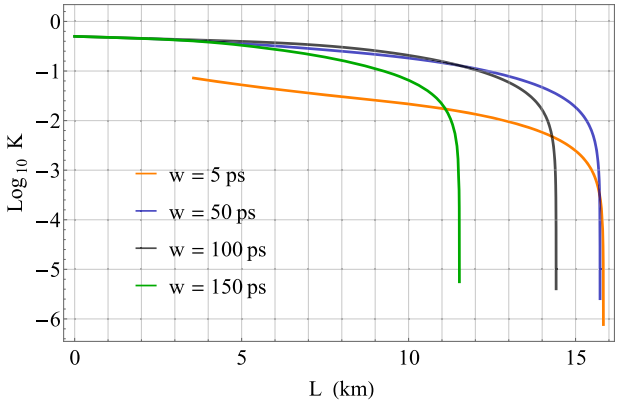

**Fig. 10** Key generation rate for four values of the detection window. Chirp parameter is fixed $C = 0$ (color figure online)

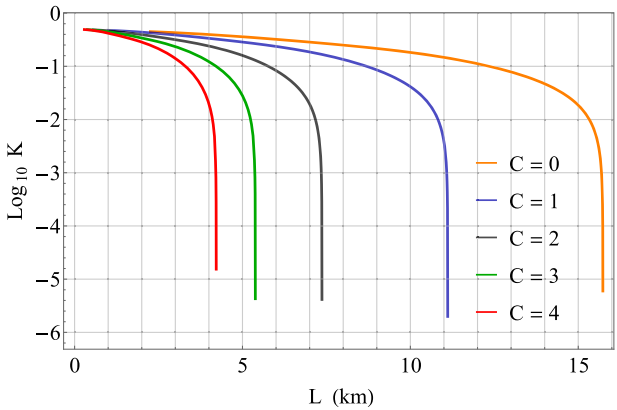

**Fig. 11** Key generation rate for non-negative values of the chirp parameter. Detection window is fixed $w = 50$ ps (color figure online)

## 5 Gaussian mode broadening

### 5.1 Methods

As a special case, let us consider a plain Gaussian mode deprived of a relative phase, In such a case, we assume that the temporal mode of a single photon can be described by a function as

$$\psi(t) = \frac{1}{\sqrt{\sqrt{2\pi}\sqrt{\sigma}}} \exp\left(-\frac{t^2}{4\sigma^2}\right), \tag{25}$$

where $\sigma$ represents the initial width of the pulse, i.e., the SD of the Gaussian distribution related to the wave packet.

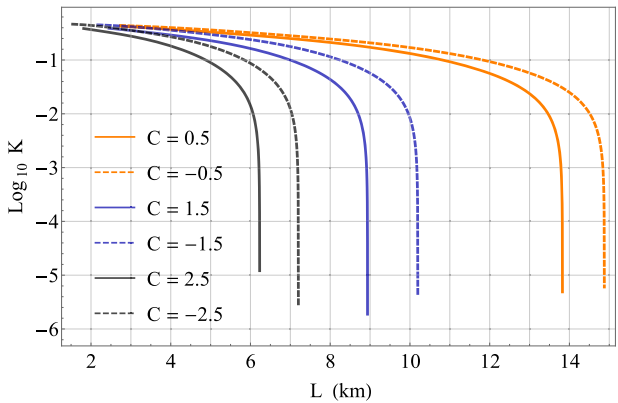

**Fig. 12** Key generation rate for opposite values of the chirp parameter. Detection window is fixed $w = 50$ ps (color figure online)

The action of the propagator (3) provides the following results

$$\psi_L(t) = \frac{(1-i)\exp\left(-\frac{t^2}{4iL\beta + 4\sigma^2}\right)}{2^{3/4}\pi^{1/4}\sqrt{\frac{L\beta}{\sigma} - i\sigma}} \quad \text{for} \quad \beta > 0 \qquad (26)$$

and

$$\psi_L(t) = \frac{(-2\pi)^{-1/4}\exp\left(-\frac{t^2}{4iL\beta + 4\sigma^2}\right)}{\sqrt{-\frac{L\beta}{\sigma} + i\sigma}} \quad \text{for} \quad \beta < 0. \qquad (27)$$

The formulas (26) and (27) represent the TWF after propagation through a medium that features normal and anomalous dispersion regimes, respectively.

When we compute the SD of the PDF after propagation, we obtain one result irrespective of the dispersion regime

$$\sigma_L = \frac{\sqrt{L^2\beta^2 + \sigma^4}}{\sigma}. \qquad (28)$$

Finally, the relative ratio $\sigma_L/\sigma$ to quantify the broadening can be obtained as

$$\Gamma = \frac{\sqrt{L^2\beta^2 + \sigma^4}}{\sigma^2} = \sqrt{1 + \frac{L^2\beta^2}{\sigma^4}}, \qquad (29)$$

which agrees with Eq. (11) for $C = 0$. This result allows us to study how the temporal shape of a single photon changes as a consequence of chromatic dispersion.

## 5.2 Results and analysis

When a photon with a plain Gaussian mode (25) is transmitted through a dispersive medium, it broadens in the time domain. The degree of broadening can be quantified by the relative ratio given by (29). $\Gamma$ can be treated as a function of $L$, which is the independent variable. Moreover, the broadening ratio $\Gamma$ depends on the GVD parameter. According to (29), for higher values of $\beta$, the degree of broadening is more significant (for both positive and negative GVD parameters).

In Fig. 13, the broadening parameter, $\Gamma$, is plotted versus the transmission distance, $L$. In the case of transmitting a signal through the air, we see that the broadening is not significant for moderate values of $L$. When $L = 200$ km, the final pulse is approx. 2.5 % wider than the initial one. The effects of chromatic dispersion are more substantial if we increase the transmission distance. In Fig. 14, we present $\Gamma$ versus $L$ in the logarithmic scale.

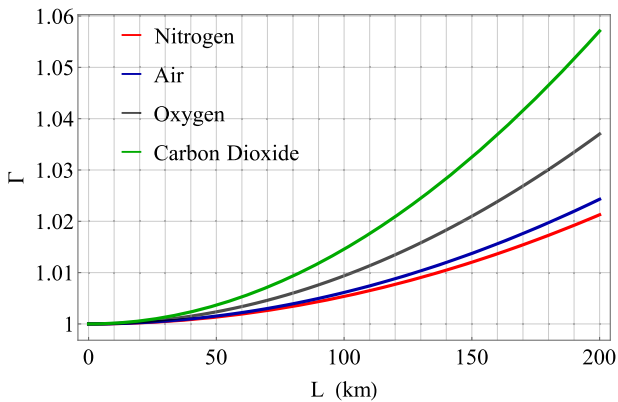

**Fig. 13** The broadening parameter, $\Gamma$, for $L \in [0, 200]$ km (color figure online)

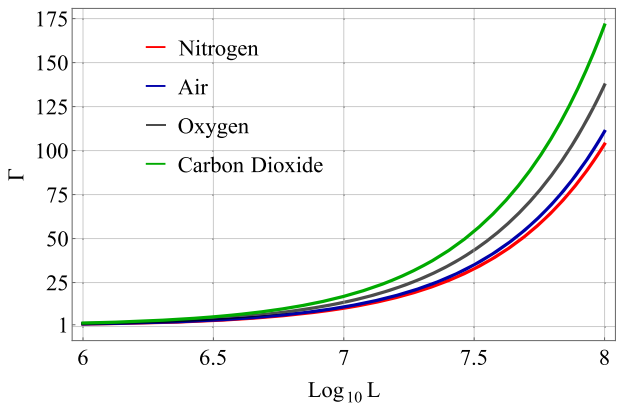

**Fig. 14** The broadening parameter, $\Gamma$, for $L \in [10^6, 10^8]$ m. A logarithmic scale was used on the horizontal axis (color figure online)

To conclude, we notice that for short and moderate values of the transmission distance, the broadening is not significant for the selected gases. However, the broadening increases with $L$, meaning that at a longer propagation distance, the broadening is more critical, which can be observed more clearly on a logarithmic scale. By following this method, one can compute $\Gamma$ according to Eq. (29) to precisely evaluate the impact of chromatic dispersion under any circumstances (for a specific transmission distance and GVD parameter).

### 5.3 Applications in quantum communications

For Gaussian wave packets, we experience only broadening irrespective of the sign of $\beta$. The fact that the photon's temporal mode is stretched negatively affects the symbol rate in quantum communications. For Gaussian modes, the symbol rate reads

$$f_S = \left(6\sqrt{\left(\frac{L\beta}{\sigma}\right)^2 + \sigma^2}\right)^{-1}. \tag{30}$$

In Fig. 15, we present the results, where the logarithmic scale has been used for better clarity. We consider two values of the GVD parameter – one corresponding to the air (given in Table 1) and the other related to a typical SMF, i.e., $\beta = -1.15 \times 10^{-26}\,\text{s}^2/\text{m}$. From Fig. 15, we see that initially, the symbol rate approximates $f_S \approx 40$ GBd. At some points, both plots start to decline, but for the air the drop-off occurs later and the curve is above the fiber.

The above analysis can determine how beneficial it is to establish an FSO channel and perform quantum communications. While FSO techniques may allow a higher symbol rate, one should bear in mind the significant limitations of such links, for example, pointing errors and turbulence-induced fluctuations. Furthermore, since we do not take into account any photon loss, the results can be treated as a theoretical limit, namely an upper bound for the symbol rate.

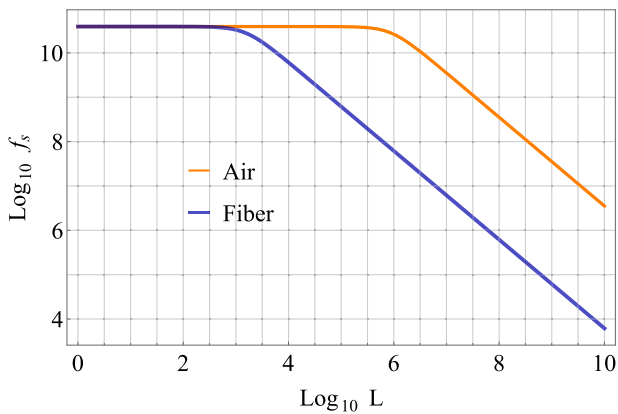

**Fig. 15** Symbol rate, $f_S$, for transmission of single photons with Gaussian modes in an SMF-28e+ fiber and through the air. The logarithmic scale is used for $L$ expressed in meters (color figure online)

## 6 Effects of chromatic dispersion on time-bin encoded qubits

### 6.1 Qubit encoded in the time domain

Time-bin qubit can be defined as a photon delocalized in the time domain in two wave packets separated by an interval $T$ [32]. Such a state can be produced by using a Mach-Zehnder interferometer to introduce a difference in the optical path. In addition, let us assume that both wave packets are Gaussian and gain an initial chirp parameter, which leads to a time-bin qubit in the form

$$\psi(t) = a\,\varphi\left(t - \frac{T}{2}\right) + b\,\varphi\left(t + \frac{T}{2}\right), \tag{31}$$

where

$$\varphi(t) = \frac{1}{\sqrt{\sqrt{\pi}\sqrt{\tilde{\sigma}}}} \exp\left(-\frac{1+iC}{2\tilde{\sigma}^2}t^2\right). \tag{32}$$

The coefficients $a$ and $b$ from Eq. (31) satisfy

$$|a|^2 + |b|^2 = 1, \tag{33}$$

which means that one can substitute $a = \cos(\theta/2)$ and $b = \sin(\theta/2)\,e^{i\phi}$, where $\theta \in [0, \pi]$ and $\phi \in [0, 2\pi)$. In addition, One can notice that the symbol $\tilde{\sigma}$ relates to the width of one wave packet.

For the wave function (31), we define the PDF in a standard way $p(t) := \psi^*(t)\,\psi(t)$. One can compute

$$\int_{-\infty}^{\infty} p(t)\,\mathrm{d}t = 1 + \sin\theta\,\cos\phi\,\exp\left(-\frac{1+C^2}{4\tilde{\sigma}^2}T^2\right). \tag{34}$$

The result in Eq. (34) indicates that there is a non-zero overlap between the basis functions. However, the PDF can be roughly normalized by adjusting the ratio $T^2/\tilde{\sigma}^2$. Since $T$ can be easily controlled in the laboratory by regulating the length of the optical path, one can ensure that for well-separated pulses and a given chirp parameter, we have $\exp\left\{-(1+C^2)T^2/(4\tilde{\sigma}^2)\right\} \approx 0$. Figure 16 presents the PDF for a qubit composed of two well-separated wave packets.

After propagation through a dispersive medium, the qubit is represented as

$$\psi_L(t) = a\,\varphi_L\left(t - \frac{T}{2}\right) + b\,\varphi_L\left(t + \frac{T}{2}\right), \tag{35}$$

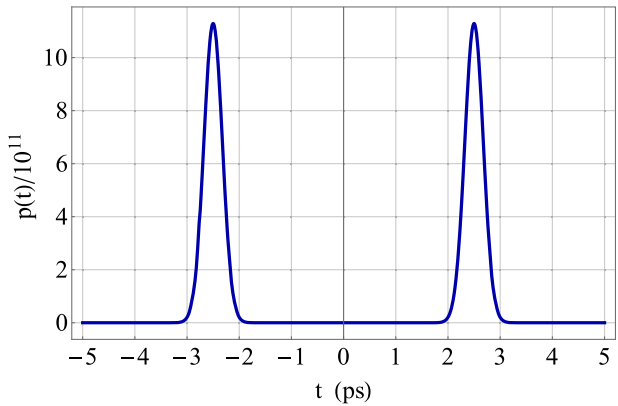

**Fig. 16** The PDF, $p(t)$, for a time-bin qubit before propagation through a dispersive medium. The parameters were: $T = 5$ ps, $\tilde{\sigma} = 0.25$ ps, $C = 2$, $\theta = \pi/2$, and $\phi = \pi/4$

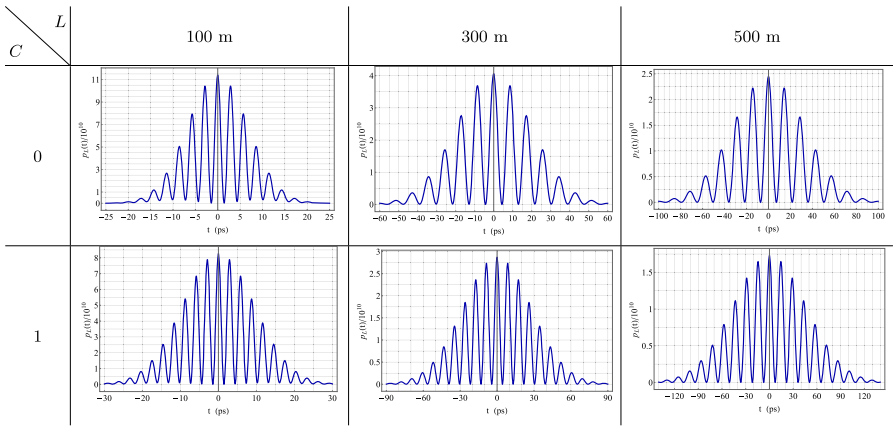

**Fig. 17** Six PDFs, $p_L(t)$, for a time-bin qubit after propagation through a dispersive medium, for different fiber lengths $L$ and phases $C$. The parameters were: $\beta = -1.15 \times 10^{-26}$ s$^2$/m, $T = 5$ ps, $\tilde{\sigma} = 0.25$ ps, $\theta = \pi/2$, and $\phi = 0$

where $\varphi_L(t) := \int_{-\infty}^{\infty} \mathcal{S}(t, \tau, L)\, \varphi(\tau)\, d\tau$. Figure 17 shows, for specific values of the parameters, six PDFs after propagation defined as $p_L(t) := \psi_L^*(t)\, \psi_L(t)$. We take into account three fiber lengths $L$ and two values of the phase $C$ (zero and non-zero) to demonstrate how both quantities affect the temporal mode of the qubit.

From the top panel of Fig. 17, when $C$ is zero, one can see how the PDF, $p_L(t)$, for a time-bin qubit after propagation through a dispersive medium changes as we increase $L$. Indeed, with the increase of $L$ from 100 to 500 m, the amplitude of $p_L(t)$ decreases, but it is more stretched over time. In comparison, when we take a non-zero $C$, the number of fringes increases, as seen from the bottom panel of Fig. 17. Therefore, the non-zero chirp parameter has a significant effect on $p_L(t)$ since it affects it twofold – by changing the number of fringes and increasing the temporal length of the qubit.

## 6.2 Results and analysis

Due to the mathematical complexity of (35), it is not possible to compute analytically the variance of the $p_L(t)$. Therefore, we proceed to this task numerically by discretization of the independent variable $L$. For fixed values of all parameters and variables, we obtain explicit one-variable functions: $p(t)$ and $p_L(t)$, for which we can compute the corresponding SDs, i.e., $\sigma$, $\sigma_L$, and then the broadening parameter as $\Gamma = \sigma_L/\sigma$. By changing $L$ step by step, we generate dotted plots that present the properties of the broadening parameter $\Gamma$.

Figure 18 shows the results for the propagation through the air while Fig. 19 includes the findings for a standard fiber. In both cases, all the parameters characterizing the qubit were fixed, i.e., $T = 5$ ps, $\tilde{\sigma} = 0.25$ ps, $\theta = \pi/2$, and $\phi = 0$ (the same as in Fig. 17).

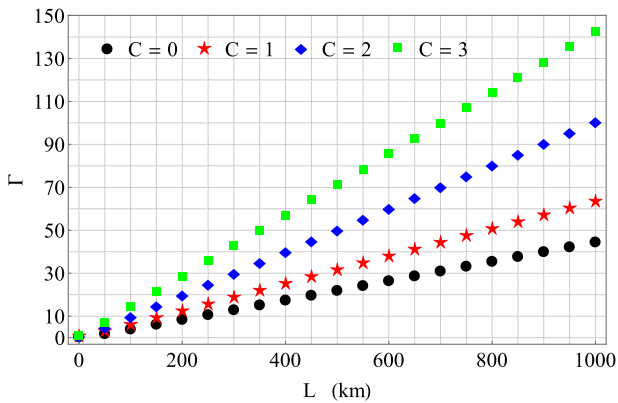

**Fig. 18** Qubit's broadening parameter, $\Gamma$, for $L \in [0, 1000]$ km and four values of $C$. The transmission through the air is considered (color figure online)

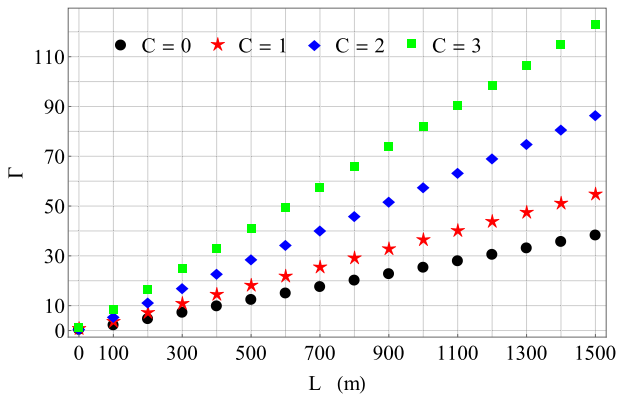

**Fig. 19** Qubit's broadening parameter, $\Gamma$, for $L \in [0, 1500]$ m and four values of $C$. The transmission through an SMF28e+ fiber is considered (color figure online)

In Fig. 18, we plot the broadening parameter $\Gamma$ versus propagation length $L$ for four values of $C$ when the transmission through the air is considered. We witness that $\Gamma$ is linear and the angle grows as the value of the chirp parameter increases from 0 to 3. By comparing Figs. 6 and 18, it can be seen that the effects of the chirp parameter on the broadening of two wave packets after propagation given by Eqs. (8) and (35) are completely different. In addition, the value of qubit's broadening in Fig. 18 is significant for long transmission lengths compared to Fig. 6.

When $C\beta \leq 0$, we illustrate the qubit's broadening as a function of $L$ during the transmission through an SMF28e+ fiber in Fig. 19. The qualitative behavior of $\Gamma$ for distinct values of the chirp parameter in this figure is similar to Fig. 18, but the quantitative behavior is different due to more significant dispersion in the fiber than in the air. Hence, the value of $\Gamma$ can be remarkably controlled by adjusting the chirp parameter to the transmission medium.

## 7 Broadening of hyperbolic-secant modes

### 7.1 Methods

Let us consider a temporal mode of a single photon that is hyperbolic-secant and can be expressed by a TWF as

$$\psi(t) = \sqrt{\frac{1}{2\sigma}\operatorname{Sech}\frac{\pi t}{2\sigma}} \exp\left(-\frac{iCt^2}{4\sigma^2}\right). \quad (36)$$

The wave function changes due to dispersion in terms of the propagator (3) acting on the initial TWF (36). It can be expressed as

$$\psi_L(t) = \frac{\exp\left(\frac{it^2}{4\beta L}\right)}{\sqrt{8\pi i \beta L \sigma}} \int_{-\infty}^{\infty} \sqrt{\frac{2}{e^{\pi\tau/(2\sigma)} + e^{-\pi\tau/(2\sigma)}}} \\ \times \exp\left\{-i\left(\frac{t}{2\beta L}\tau + \left(\frac{1}{4\beta L} - \frac{C}{4\sigma^2}\right)\tau^2\right)\right\} d\tau. \quad (37)$$

The broadening parameter $\Gamma$ of the hyperbolic-secant modes cannot be computed analytically. Similarly to the cases of generalized Gaussian modes, we can obtain $\Gamma$ numerically for the specific parameters characterizing hyperbolic-secant modes.

### 7.2 Results and analysis

The results presented in this section focus on the chirp parameter as a crucial factor. We also distinguish two scenarios – positive and negative dispersions.

In Fig. 20, we present the plots for positive dispersion and non-negative chirp parameters. The dependence of the broadening parameter on the chirp parameter suggests that careful adjustment of the chirp can effectively mitigate the detrimental effects of normal dispersion. The presence of a minimum in broadening indicates an optimal

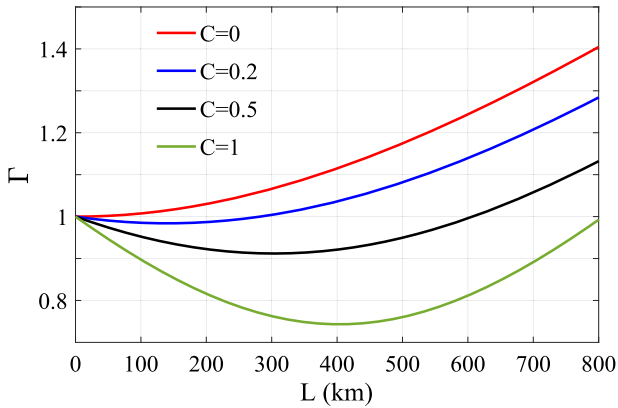

**Fig. 20** The broadening parameter, $\Gamma$, for $L \in [0, 800]$ km, $\sigma = 4.25$ ps and four non-negative values of $C$ for the hyperbolic-secant pulses. The transmission through the air is considered (color figure online)

chirp value for minimizing temporal spreading. The deeper minimum for higher chirp values suggests a potential optimization strategy. By increasing the chirp, one can achieve a more pronounced reduction in broadening before the subsequent increase. This insight is valuable for practical implementations, allowing for more efficient compensation of dispersion effects.

The monotonic increase in broadening for non-positive chirps highlights a contrasting behavior, as presented in Fig. 21. This could indicate the challenges associated with compensating for dispersion using certain chirp values, emphasizing the importance of positive chirps for effective dispersion management in FSO communications.

The reversal of trends for negative dispersion is a noteworthy observation. For non-negative chirp parameters, we observe a monotonic increase of $\Gamma$, which is provided in Fig. 22 for the transmission through an SMF28e+ fiber. The initial decline

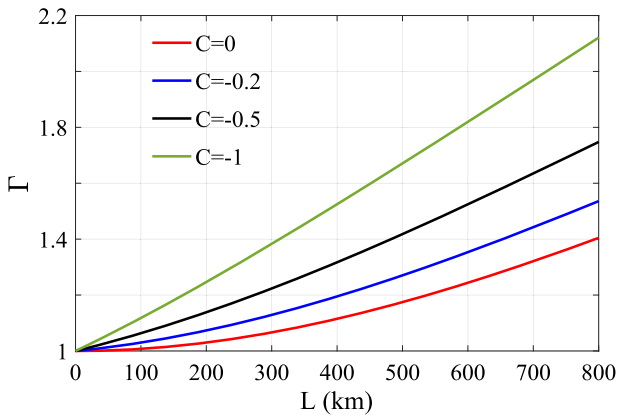

**Fig. 21** The broadening parameter, $\Gamma$, for $L \in [0, 800]$ km, $\sigma = 4.25$ ps and four non-positive values of $C$ for the hyperbolic-secant pulses. The transmission through the air is considered (color figure online)

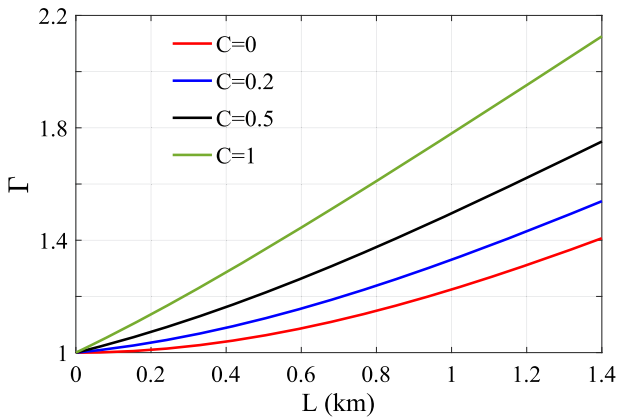

**Fig. 22** The broadening parameter, $\Gamma$, for $L \in [0, 1.4]$ km, $\sigma = 4.25$ ps and four non-negative values of $C$ for the hyperbolic-secant pulses. The transmission through an SMF-28e+ fiber is considered (color figure online)

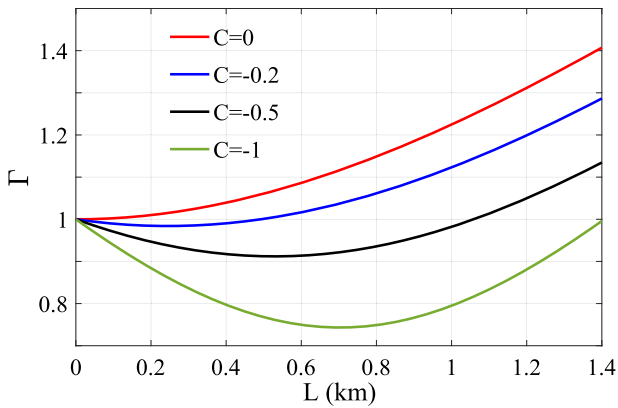

**Fig. 23** The broadening parameter, $\Gamma$, for $L \in [0, 1.4]$ km, $\sigma = 4.25$ ps and four non-positive values of $C$ for the hyperbolic-secant pulses. The transmission through an SMF-28e+ fiber is considered (color figure online)

in broadening for negative chirps, as given in Fig. 23, suggests a compensatory effect, where selected chirp values counteract the broadening caused by negative dispersion. Tailoring the chirp parameter in the presence of negative dispersion can potentially enhance communication rates over specific distances, offering a novel avenue for system optimization.

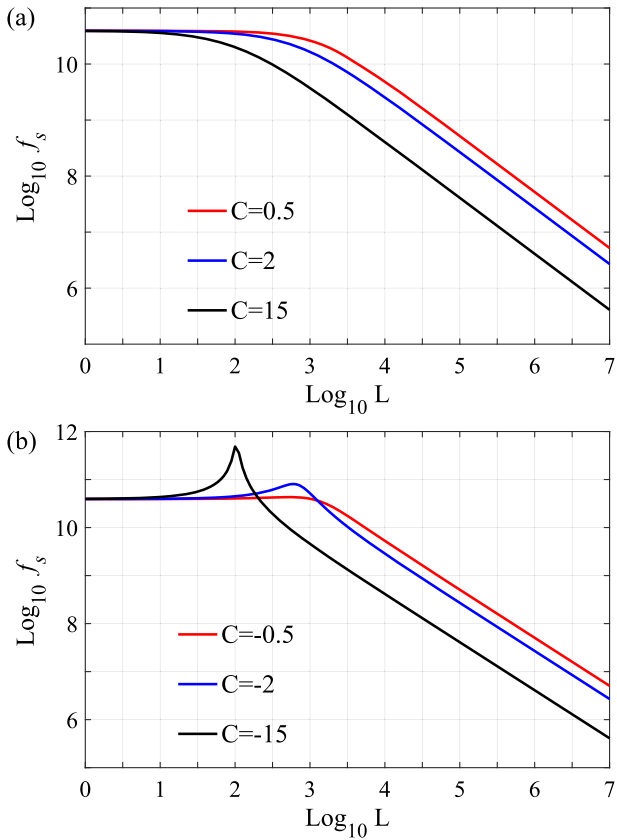

**Fig. 24** Symbol rate, $f_S$, for transmission of photons with hyperbolic-secant pulses with **a** positive chirp parameters and **b** negative chirp parameters through an SMF28e+ fiber. Logarithmic scale is used for $L$ expressed in meters (color figure online)

### 7.3 Applications in quantum communications

To investigate the efficiency of hyperbolic-secant modes in fiber-based quantum communications, we compute the symbol rate for negative dispersion, as shown in Fig. 24. The consistent decline in symbol rate with positive chirps underlines the challenging nature of maintaining high communication rates in the presence of anomalous dispersion. These findings emphasize the need for compensatory techniques to counteract the adverse effects of dispersion on communication efficiency.

The observed pitch in the symbol rate for negative chirps introduces a valuable contribution. The ability to tailor the chirp parameter to exploit dispersion for increased communication rates over specific distances presents a strategic advantage. This result opens up possibilities for optimizing quantum communication systems in scenarios involving negative dispersion.

# 8 Discussion and outlook

Our research focuses on the effects of chromatic dispersion on single-photon TWFs in the context quantum communications, introducing a paradigm shift from foregoing studies oriented on classical beams. Working within the quantum regime, we analyze generalized Gaussian modes, establishing a foundation for the exploration of chirped and unchirped Gaussian TWFs. The key aspect lies in our emphasis on the quantum context, differentiating our work from existing literature that predominantly addresses classical beam spreading.

We demonstrate how various parameters influence the spreading of TWFs and, consequently, the maximum symbol rate in quantum communications. This comprehensive approach sets our work apart, providing a deeper understanding that goes beyond the models presented in the existing literature.

Temporal characteristics of single photons must be investigated to optimize the performance of QKD schemes [40]. Our research contributes to these efforts by demonstrating the impact of the chirp parameter on the key generation rate in the BB84 protocol.

Notably, our study also involves the analysis of time-bin qubits under the influence of chromatic dispersion, reflecting the current trajectory of ongoing research in quantum information processing [41]. This perspective enhances the relevance and applicability of our findings in the rapidly evolving field of quantum communications.

While our primary investigation focuses on picosecond pulses, the principles and findings presented in this study can be extended to femtosecond and nanosecond pulses. Femtosecond pulses, due to their shorter duration and broader spectral bandwidth, experience more pronounced effects of chromatic dispersion. The temporal broadening caused by dispersion is significantly greater for femtosecond pulses, making the chirp parameter's role in dispersion management even more critical. Precise control of the chirp parameter can effectively mitigate dispersion effects, ensuring high-fidelity transmission of quantum information. Conversely, nanosecond pulses, with their longer durations and narrower spectral bandwidths, are less susceptible to chromatic dispersion. Although the impact is reduced, chirp management techniques can still be applied to compensate for any residual dispersion effects, optimizing the performance of nanosecond pulses in quantum communication systems.

Our results contribute not only to the theoretical foundations of quantum communication but also offer practical insights for system designers. By highlighting the quantum nature of dispersion effects and their impact on time-bin qubits, our research addresses a critical gap in the current body of knowledge, paving the way for advancements in the design and optimization of quantum communication systems.

**Acknowledgements** A.C. acknowledges financial support from the Foundation for Polish Science within the "Quantum Optical Technologies" project carried out within the International Research Agendas programme co-financed by the European Union under the European Regional Development Fund (MAB/2018/4). X.C. acknowledges support from the National Natural Science Foundation of China (Grant No. 12005121). S.H. was supported by Semnan University under Contract No. 21270.

**Data availability** No datasets were generated or analyzed during the current study.

## Declarations

**Conflict of interest**  The authors declare that they have no known competing financial interests.